\documentclass[10pt,twocolumn,letterpaper]{article}

\usepackage{cvpr}              %

\usepackage{algorithm}
\usepackage{algorithmic}

\usepackage{multirow}
\usepackage{makecell}
\definecolor{cvprblue}{rgb}{0.21,0.49,0.74}
\usepackage{graphicx}
\usepackage[pagebackref,breaklinks,colorlinks,allcolors=cvprblue]{hyperref}

\title{Sonic: Shifting Focus to Global Audio Perception in Portrait Animation}

\author{
\textbf{Xiaozhong Ji}\textsuperscript{1*}
\ 
\textbf{Xiaobin Hu}\textsuperscript{1*}
\ 
\textbf{Zhihong Xu}\textsuperscript{2}
\ 
\textbf{Junwei Zhu}\textsuperscript{1}
\ 
\textbf{Chuming Lin}\textsuperscript{1} 
\
\textbf{Qingdong He}\textsuperscript{1} 
\\
\textbf{Jiangning Zhang}\textsuperscript{1} 
\ 
\textbf{Donghao Luo}\textsuperscript{1} 
\ 
\textbf{Yi Chen}\textsuperscript{1} 
\ 
\textbf{Qin Lin}\textsuperscript{1}
\
\textbf{Qinglin Lu}\textsuperscript{1} 
\ 
\textbf{Chengjie Wang}\textsuperscript{1}
\\
\textsuperscript{1} Tencent
\quad
\textsuperscript{2} Zhejiang University
}

\begin{document}

\twocolumn[{%
\renewcommand\twocolumn[1][]{#1}%
\maketitle
\begin{center}
    \centering
    \captionsetup{type=figure}
\includegraphics[width=0.9\textwidth]{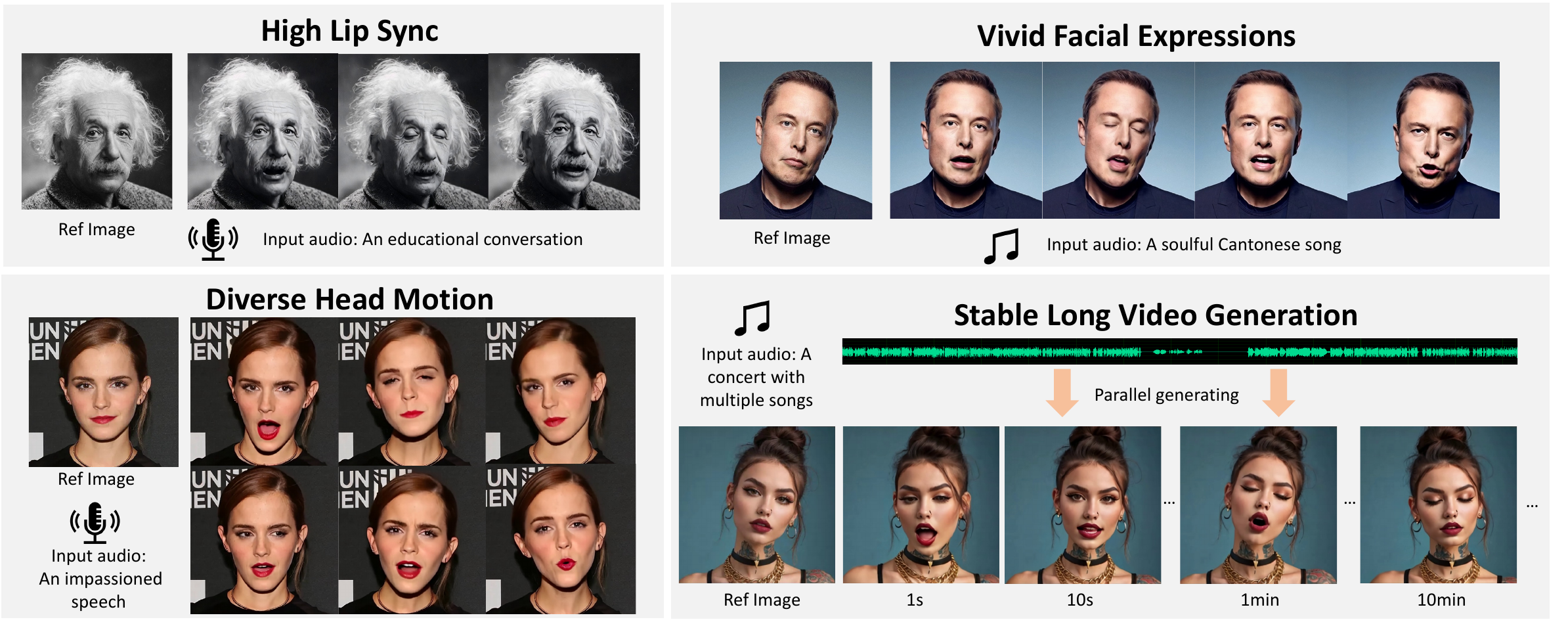}
    \vspace{-10pt}
    \captionof{figure}{\textbf{Sonic excels in producing vivid portrait animation videos given a reference image and an audio clip.} Beyond its fundamental lip-syncing capabilities, Sonic demonstrates proficiency in creating a diverse spectrum of facial expressions and adaptable head movements. Notably, when dealing with extented videos, Sonic can yield stable and seamless outcomes in a parallel fashion, all while maintaining an unified paradigm focusing on global audio perception. 
    }
    \label{fig:teaser}
\end{center}%
}]

\if TT\insert\footins{\noindent\footnotesize{
*Equal Contribution  \\
Project page: \url{https://jixiaozhong.github.io/Sonic/}.}}\fi

\begin{abstract}
The study of talking face generation mainly explores the intricacies of synchronizing facial movements and crafting  visually appealing, temporally-coherent animations. 
However, due to the limited exploration of global audio perception, current approaches predominantly employ auxiliary visual and spatial knowledge to stabilize the movements, which often results in the deterioration of the naturalness and temporal inconsistencies.
Considering the essence of audio-driven animation, 
the audio signal serves as the ideal and unique priors to adjust facial expressions and lip movements, without resorting to interference of any visual signals. 
Based on this motivation, we propose a novel paradigm, dubbed as Sonic, to {s}hift f{o}cus on the exploration of global audio per{c}ept{i}o{n}.
To effectively leverage global audio knowledge, we disentangle it into intra- and inter-clip audio perception and collaborate with both aspects to enhance overall perception.
For the intra-clip audio perception, 1). \textbf{Context-enhanced audio learning}, in which long-range intra-clip temporal audio knowledge is extracted to provide facial expression and lip  motion priors implicitly expressed as the tone and speed of speech. 2). \textbf{Motion-decoupled controller}, in which the motion of the head and expression movement are disentangled and independently controlled by intra-audio clips. Most importantly, for inter-clip audio perception, as a bridge to connect the intra-clips to achieve the global perception, \textbf{Time-aware position shift fusion}, in which the global inter-clip audio information is considered and fused for long-audio inference via through consecutively time-aware shifted windows. 
Extensive experiments demonstrate that the novel audio-driven paradigm outperform existing SOTA methodologies in terms of video quality, temporally consistency, lip synchronization precision, and motion diversity.

\end{abstract}

\section{Introduction}
\label{sec:intro}
Talking Face Animation is a technique aimed at creating a lifelike representation of a speaking person by animating a still image in sync with speech audio.
This technology holds great potential in enhancing the realism of virtual reality \cite{bozkurt2023speculative}, improving the efficiency of film and gaming making \cite{kal2024educational}, facilitating accessibility for individuals with communication impairments \cite{johnson2018assessing}, and offering therapeutic support and social interaction in healthcare \cite{rehm2016role}. 
Prominent progress in this domain is demonstrated by notable works based on Stable Diffusion \cite{rombach2022high}, which project the parametric or implicit representations of lip movements \cite{prajwal2020lip, zhang2023sadtalker}, 
facial expressions \cite{he2023gaia, ma2023dreamtalk, shen2023difftalk}, 
and head movement \cite{sun2023vividtalk} 
in the latent space \cite{corona2024vlogger, liu2024anitalker, tian2024emo, xu2024vasa}
to generate high-quality video with a high level of realism and liveliness.

However, precise audio control and temporal coherence have not yet been effectively achieved. Current approaches \cite{tian2024emo, xu2024hallo, chen2024echomimic, zhang2024lingualinker, wei2024aniportrait, wang2024v, zhang2023sadtalker, jiang2024loopy} predominantly addressed these two issues in a completely segregated manner, overlooking the holistic coordination between audio and vision. For audio control, audio features were segmented according to timestamps and matched to each visual frame. By applying cross-attention within a spatial latent space, each frame was restricted to the nearby audio information, which hindered the transformation into an optimal motion representation. To achieve temporal coherence, self-attention along the temporal dimension was applied to smooth visual features within a clip, while overlapping or motion frames were used to enhance coherence between different clips. However, these stability strategies have limited temporal receptive fields, which can reduce the richness of motion. Additionally, they do not take audio information into account, potentially disrupting the synchronization between audio and visual elements.

To address these issues,  we propose a novel paradigm, dubbed as Sonic, to extent the range of receptive field to the global level guided by the audio priors. Sonic aims to exploit global audio perception rather than the motion frames and other visual motion. Performing as long-range and global signal, audio shows tone, rhythm  and speed of speech which implicitly provides expression and head movement motion priors.  
However, we have to face such a tricky challenge that, as a weak correlation cross-modality signal without the guidance of any visual motion (\textit{e.g.,} motion frames), current audio learning paradigm leads to the temporal jitter and unsatisfactory pattern and fails to well-align the audio and motion. 
To this end, we first customize efficient global audio perception learning mechanism. Specifically, \textbf{context-enhanced audio learning} is proposed as long-range audio learning module to extract the intra-clip audio temporal knowledge from the input audio clip. Then such intra-clip audio temporal information is mapped as a temporal-embeddings for the latter fusion by audio-temporal cross-attention. The temporal-embeddings projection operation can effectively reduce the computational burden and capture temporal priors within current audio clip. 
Although context-enhanced audio learning directs the temporal movements correlated with audio, such as lip movements synchronized with speech and facial expressions related to the content, there still exists the absence of habitual head movements.
Secondly, we design \textbf{motion-decoupled controller} to disentangle the habitual head and expression movement and support the independent control via two explicit parameters learned from the current intra-clip audio. To enrich the functionality and playability of animation, we also add an plug to allow users to define some exaggerated movements.
Most importantly, for inter-clip audio perception, 
\textbf{time-aware position shift fusion} enlarges the intra-clip audio perception to the global inter-clip audio perception via time-aware shifted windows consecutively bridging the preceding clip, providing the global inter-clip  connection that significantly enhance modeling temporal power, as demonstrated in Figure~\ref{fig:teaser}. It is worth noting that our time-aware position shift fusion requires no extra training cost compared with the motion frames in most recent works, such as EMO~\cite{tian2024emo}, Loopy \cite{jiang2024loopy} and also no additional inference time caused by overlapping frames.
Our contributions are summarized as:
\begin{itemize}
    \item We propose a novel unified paradigm without the guidance of visual motion, dubbed as Sonic, to focus on the exploration of global audio perception.
    \item 
    For intra-clip audio perception, context-enhanced audio learning and motion-decoupled controller excavate the intra-clip audio temporal knowledge to direct 
    the temporal movements. 

    \item For inter-clip audio perception, we propose time-aware position shift fusion to enhance the global inter-clip audio perception by extending the intra-clip audio perception through consecutively time-aware shifted windows.  
    
    \item Extensive experiments demonstrate that the novel unified global audio perception paradigm outperforms existing  SOTA methods in terms of video quality, temporal consistency, lip synchronization, and motion diversity.
\end{itemize}

\section{Related Works}
\label{sec:formatting}
\subsection{Diffusion-based Video Generation}
The remarkable achievements in diffusion models in image generation have sparked widespread research interest for video generation. Early endeavors in video generation \cite{esser2023structure, ho2022video, hong2022cogvideo, khachatryan2023text2video, singer2022make}
have primarily concentrated on devising image-based diffusion models with temporal module to effectively capture the dynamic characteristics inherent in video sequences. To leverage the powerful ability of pretrained image diffusion models,  Video LDM \cite{blattmann2023align} incorporates temporal layers that learn to align images in a temporally consistent manner.  
AnimateDiff \cite{guo2023animatediff} provides a motion-specific module by multi-stage progressive training manner, and the motion module can be seamlessly integrated into existing text-to-image architectures without any further modifications. 
VideoComposer \cite{wang2024videocomposer} and VideoCrafter \cite{chen2024videocrafter2} further explore the synthesis of image-to-video generation via  textual and visual features combinations. Despite these notable advancements, the field still struggles to overcome the challenge of generating temporal coherent and high-quality videos. As a powerful motion representation trained on a large well-curated high-quality dataset, Stable Video Diffusion \cite{blattmann2023stable} shows the remarkable temporal-consistency and the high-quality visual characters for downstream tasks. There exists a large gap between the audio-driven portrait animation and general video generation due to the fact that audio-driven task aims to use the weak cross-modality audio signal to precisely control motion movements (\textit{e.g.,} lip and expression movement).

\subsection{Audio-driven Talking Face Generation}
The generation of talking face videos from audio inputs has been a longstanding task. In the early stages, the emphasis was primarily on synthesizing lip movements alone, accomplished by directly mapping audio signals to lip movements while keeping other facial attributes unchanged \cite{suwajanakorn2017synthesizing, ji2024realtalk, chen2018lip, prajwal2020lip, yin2022styleheat}.
In more recent efforts, the scope has been broadened to encompass a wider range of facial expressions and head movements, which are derived from audio inputs and a single reference image. These methods can be broadly classified into with or without intermediate representation. 
For instance, by employing 3D coefficients representation (\textit{e.g.,} Sadtalker \cite{zhang2023sadtalker}, Dreamtalk \cite{ma2023dreamtalk}, AniPortrait \cite{wei2024aniportrait}, VASA-1 \cite{xu2024vasa}), the video generation process can be informed through the direct or indirect visible control signal in the aspect of head movement and expression. 
Nevertheless, a recurring challenge faced by these techniques is the limited capacity of the 3D mesh to capture intricate details, consequently constraining the overall dynamism and realism. In contrast, the methods that omit intermediate representation exhibit superior naturalness and consistent preservation of identity with the original image. 
EMO \cite{tian2024emo}, Hallo \cite{xu2024hallo}, and Loopy \cite{jiang2024loopy} employ a direct audio-to-video synthesis approach to ensure a high level of realism and naturalness, eliminating the need for intermediate 3D representation. However, these approaches primarily rely on 
additional conditions associated with spatial motion or motion frames to maintain temporal consistency.
In this regard, they overlook the crucial role of audio in audio-driven animation, consequently diminishing both expressiveness and audio-motion synchronization.

\subsection{Long Video Inference}
Significant research efforts have been devoted to extending the duration of generated videos, aiming to expand their practical applications. Autoregressively predicting successive frames \cite{he2022latent, voleti2022mcvd}
and hierarchical coarse-to-fine methods \cite{yin2023nuwa}
have emerged to generate long videos. Then Lumiere \cite{bar2024lumiere} divides the video into overlapping temporal segments, denoises each segment independently, and finally fuse these segments in an optimization algorithm. 
However there is no available high-efficiency method for audio-condition global perception.
As one step towards achieving stronger capabilities, our Sonic proposes time-aware position shift fusion strategy progressively denoise intra-clip audio to establish a global inter-clip connection through the use of time-aware shifted windows that bridge the preceding clip along the timesteps axis. 
This inter-clip fusion with global receptive field needs no additional training cost and introduce no extra inference time without overlapping frames.

\section{Methodology}
\begin{figure*}[ht!]
    \centering
    \includegraphics[width=0.9\linewidth]{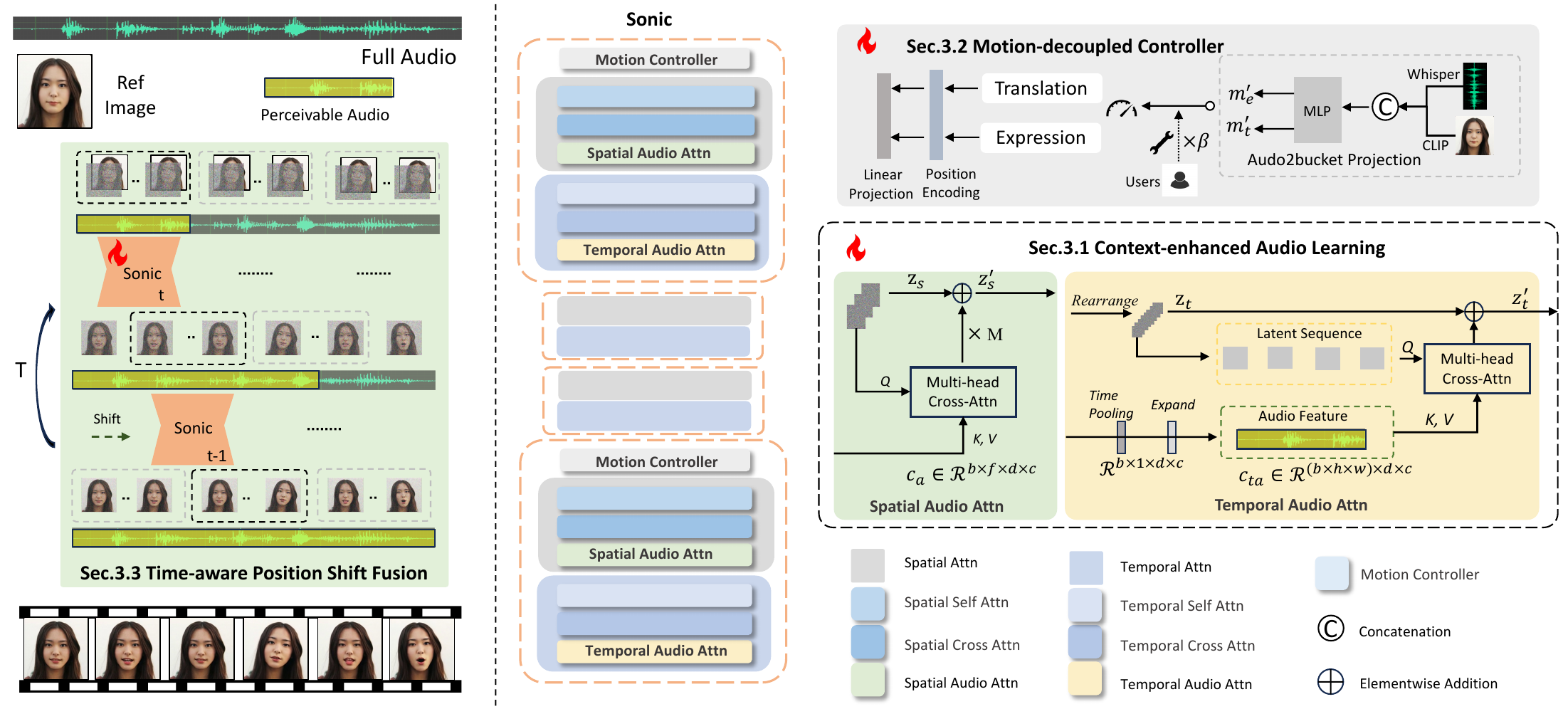} %
    \vspace{-4mm}
    \caption{\textbf{Framework of our approach.} 
    Sonic processes each clip of the long audio in parallel, shifting to a new context at each time step to progressively fuse inter-clip latent features across global audio perception. In Sonic, we enhance intra-clip temporal audio context learning and decouple motion to improve dynamics. 
    } 
    \vspace{-3mm}
    \label{fig:framework}
\end{figure*}

\noindent \textbf{Overall framework.}
Given a single portrait reference image and an input audio, our one-stage framework can generate a portrait video well-synchronized with the driven audio.
Our proposed Sonic framework aims to focus on the exploration of global audio perception, primarily consisting of context-enhanced audio learning, motion-decoupled controller, and time-aware position shift fusion module, as illustrated in Figure~\ref{fig:framework} .

\subsection{Context-enhanced Audio Learning}
For audio-talking face animation, the audio is the ideally unique and global signal to adjust the face expression and movement without interference from other visual signals. Thus, the motivation behind context-enhanced audio learning is to guide the generation of spatial and temporal frames solely based on audio signals, rather than motion frames or other visual signals, to achieve the high-realism generation. 
Specifically, we adopt Whisper-Tiny~\cite{radford2023robust} as the pretrained audio feature extraction model, which is more lightweight than Wav2Vec~\cite{baevski2020wav2vec} used by most existing methods~\cite{tian2024emo,xu2024hallo,chen2024echomimic}. 
For each audio frame, the features from the last layers of five stages are 
concatenated together for multi-scale understanding. 
For each video frame, a duration of 0.2\textit{s} audio feature is used frame-wise to give rich context.
We use three linear layers to project audio feature to match the dimension of cross attention. 
The transformed audio embedding is denoted as $c_a \in \mathcal{R}^{b \times f\times d \times c}$, where $b$ denotes batch size, $f$ is audio length, $d$ is num of context tokens, and $c$ is hidden size of the cross-attention.
To focus on talking face area, the core latent features is restricted with a mask $M$ determined by a joint bounding boxes of faces. 
The audio signal can act as the spatial knowledge provider being injected into the spatial cross-attention layers as follows:
\begin{equation}
\small
\begin{aligned}
    z_{s}^{'} &=  z_{s} + \operatorname{CrossAttn}(Q(z_s), K(c_a), V(c_a)) \cdot M,
\end{aligned}
\end{equation} 
where $z_{s}$ is spatial latent features of self-attention layer, and $z_{s}^{'}$ is the adjusted spatial features guided by audio signals in spatial-aware level.

Furthermore, audio is a long-range and global signal that provides expression and head movement motion priors implicitly expressed through the tone and speed of speech. To excavate the temporal information from the long-range audio knowledge, we customize a temporal-audio module for precise motion control. 
The temporal module in existing methods \cite{tian2024emo,zhang2023sadtalker} follows AnimateDiff, which performs only self-attention across the temporal dimension without audio aids. 
Some other talking-face \cite{jiang2024loopy, xu2024hallo} works extract the motion priors from the visual knowledge (\textit{e.g.,} motion frames), which deviates from the essence of audio-driven task that relies solely on audio to drive movement. 
Thus we propose a temporal audio cross-attention module, introducing audio embedding to guide temporal alignment. 

The preceding latent from temporal self-attention layers can be reshaped as $z_t \in \mathcal{R}^{(b\times h \times w)\times f \times c}$. To reduce the computation burden, the audio feature $c_a$ is average-pooled  along temporal dimension as the shape $\mathcal{R}^{(b\times 1 )\times d \times c}$, then repeated and reshaped as $c_{ta} \in \mathcal{R}^{(b\times h \times w)\times d \times c}$. 
The projected audio temporal-embeddings are injected into the denoising U-Net via a temporal cross-attention blocks as:
\begin{equation}
\small
\begin{aligned}
    z_{t}^{'} &=  z_{t} + \operatorname{CrossAttn}(Q(z_t), K(c_{ta}), V(c_{ta})).
\end{aligned}
\end{equation} 
The spatial and temporal audio attention layers are cascaded in multiple down-stages and up-stages to enable natural appearance and smooth temporal movement.

\subsection{Motion-decoupled Controller}
Context-enhanced Audio Learning governs the temporal movements strongly correlated with audio, such as lip movements synchronized with speech and facial expressions related to the content. 
On the other hand, our motion-decoupled controller performs a complementary role by directing the temporal movements weakly associated with audio, such as habitual head movements and random changes in perspective. 
This controller can disentangle motion into explicit head and expression movement amplitudes, thereby enhancing both functionality, playability, and interactivity. 
Thus, we introduce independent augmented motion-bucket parameters which effectively influence motion patterns of head movements and expression strength. 
During the training phase, the translation motion-bucket $m_t$ is computed as the variance of bounding boxes of video clip, while the expression motion-bucket $m_e$ is variance of relative landmarks. 
Both buckets are integers in the range of [0, 128]. Through position encoding and linear projection, they are added into ResNet blocks as follows: 
\begin{equation}
\small
\begin{aligned}
    emb &= W[\operatorname{Pe}(m_t),\operatorname{Pe}(m_e)],\\
\end{aligned}
\end{equation} 
where $emb$ is the embedding, $\operatorname{Pe}$ is the position encoding, and $W$ represents the linear projection weights.

For practice application, motion-decoupled controller also supports the automatically adjustable parameters learning from the audio and reference image, and deeply explore the correlation between audio and portrait reference without requiring the parameter setting from users. 
To simplify multiple parameters and maintain adjustability, the predicted motion buckets are multiplying by a scale $\beta$ ($0.5$ for mild dynamic, $1.0$ for moderate dynamic, and $2.0$ for intense dynamic).
Specifically, we utilize a audio-to-bucket manner to adaptively predict the motion-buckets relying on the audio and reference image. 
In inference phase, the motion buckets are predicted and rescaled as:
\begin{equation}
\small
\begin{aligned}
    m_t^{'},m_e^{'} = \beta \times \operatorname{E_b}(c_a, R_{img}).
\end{aligned}
\end{equation}
where $m_t^{'}$, $m_e^{'}$ are the estimated translation and expression motion-bucket achieved by the audio $c_a$ and single reference image CLIP embedding $R_{img}$, and $\operatorname{E_b}$ is the audio-to-bucket projection consisting of there linear layers with the ReLU as activation function.

\subsection{Time-aware Position Shift Fusion}

\begin{figure}[ht!]
    \centering
    \includegraphics[width=0.99\linewidth]{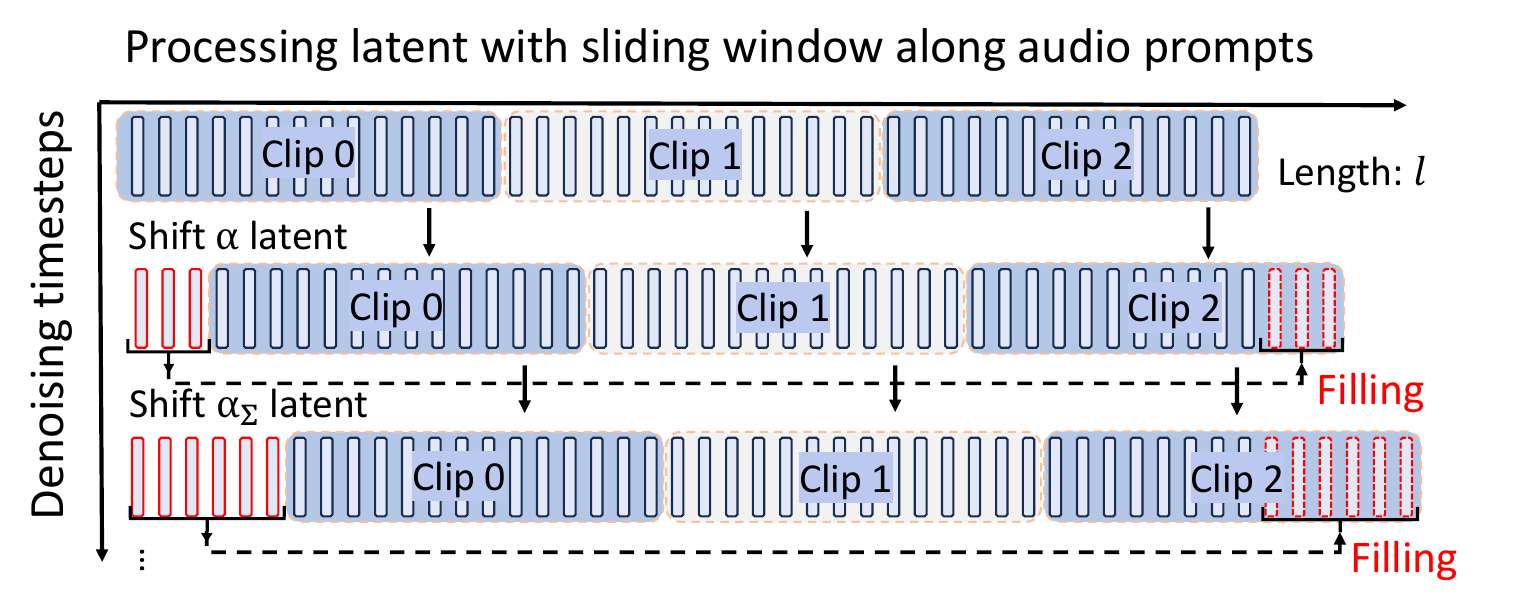}
    \vspace{-4mm}
    \caption{\textbf{Illustration of the proposed time-aware position-shift fusion.} The model processes each clip non-overlapping. In next timestep, the model starts from a new position determined by the offset, thereby integrating long-range context. Specifically, the tail latents are filled cyclically from the head.} 
    \label{fig:shift_denoise}
\end{figure}

Recent methods~\cite{tian2024emo,xu2024hallo,jiang2024loopy} use motion frames or overlapping latent during coherent denoising, which suffer from a limited receptive field and increased the training or inference computational complexity for audio-conditioned video generation problem.
In contrast, we propose a bridge, the time-aware position shift fusion, to extent the intra-clip audio perception to the global inter-clip audio-perception via consecutively time-aware shifted windows. The simple yet effective fusion strategy, without increasing extra inference or training time costs, can effectively address jitter and abrupt transition issues.
Audio serves as the input condition for the talking face, where longer audio prompts the generation of longer videos. 
Based on this principle, we tailor time-aware position-shift fusion mechanism along the timestep axis where the latent sliding window is progressively shifted by a certain length at different timesteps while also preserving the frame correspondence relationship between audio and video signal.

Specifically, as described in Figure~\ref{fig:shift_denoise} and Algorithm~\ref{alg:shift-denoise}, our time-aware position-shift fusion consist of two nested loops. The outer loop is inverse diffusion process, while the inner loop is a sliding window process that model predicts for the full-length audio conditions.  
At each timestep, the model takes a clip of audio input to predict the corresponding latent with a start offset to smooth the segment from the last timestep. The video diffusion model naturally bridges the context, generating continuous movement following the audio prompts. In detail, we set a cumulative shift offset $\alpha _{\Sigma}$, accumulating 
 $\alpha$ at each timestep to ensure that the model starts sliding window from different positions. The offset $\alpha$ is experimentally set to a small integer, such as $3,7$; we will report analysis on this parameters in experiments part. Note that the relative positioning between the latents and the audio prompts has not been altered. The only modification occurs in their recombination during the input phase to the model. In a special case, at the end of the long sequence, the shifted position index may exceed the sequence length. To address this issue, we adopt a circular padding strategy, padding the start audio prompt and latents to the end. 

\begin{algorithm}[ht!] 
\small
    \caption{Time-aware Position-shift Fusion} 
    \label{alg:shift-denoise} 
    \begin{algorithmic}[1] %
        \REQUIRE Audio embedding $c_a^{[0,l]}$ with length $l$, denoising steps $T$, initial noisy latent $z_T^{[0,l]}$, pretrained Sonic model $\operatorname{ST}(\cdot)$ for sequence length $f$, position-shift offset $\alpha < f < l$.\\ 
        \ENSURE Denoised latent $z_0^{[0,l]}$.
        \STATE Initialize accumulated shift offset $\alpha _{\Sigma}=0$. 
        \FOR{$t = T,\cdots,1$} 
        \STATE // Denoising loop 
        \STATE Initialize start point $s=\alpha _{\Sigma}$, end $e=s+f$, processed length $n=0$. // start from new position for each timestep.
        \WHILE{$n < l$} 
        \STATE // Sequence loop
        \STATE $z_{t-1}^{[s,e]} = \operatorname{ST} (z_{t}^{[s,e]}, c_a^{[s,e]}, t)$ 
        \STATE $s \xleftarrow{} s + f$,  $e \xleftarrow{} e + f$, $n \xleftarrow{} n + f$. // Move to next clip non-overlapping
        \IF{$s > l$ \OR $e > l$}
        \STATE $s \xleftarrow{} s \% l $, $e \xleftarrow{} e \% l  $. // Padding circularly
        \ENDIF
        \ENDWHILE
        \STATE $\alpha _{\Sigma} \xleftarrow{} \alpha _{\Sigma} + \alpha $. // Accumulate shift offset
        \ENDFOR
        \RETURN Denoised latent $z_0^{[0,l]}$.
    \end{algorithmic}
\end{algorithm}
\vspace{-2mm}

\begin{table*}[ht!]
    \centering
    \small
    \renewcommand\arraystretch{0.9}
    \caption{\textbf{Quantitative comparisons with the state-of-the-arts on the HDTF and CelebV-HQ test set.} $^*$ indicates GAN-based methods and others are Diffusion-based methods. The best results are \textbf{bold}, and the second are \underline{underlined}.}
    \label{tab:qualitative_experiments}
    \vspace{-2mm}
    
    \resizebox{0.8\linewidth}{!}{
    \small
\begin{tabular}{c | c | c| c| c| c| c| c| c| c}
        \toprule
        \textbf{Dataset} & \textbf{Method} & \textbf{FID$\downarrow$} & \textbf{FVD$\downarrow$} & \textbf{Sync-C$\uparrow$} & \textbf{Sync-D$\downarrow$} & \textbf{E-FID$\downarrow$} & \textbf{F-SIM$\uparrow$} & \textbf{Smooth$\uparrow$} & \textbf{Runtime$\downarrow$}  \\
        \midrule
            \multirow{7}{*}{\makecell{HDTF}} 
             &  SadTalker$^*$~(2023)~\cite{zhang2023sadtalker}   & 61.672 &  397.114 & 1.755 & 10.695
             & 2.482 & 0.9287 & 0.9961  & 3.75\\
             &  Aniportrait~(2024)~\cite{wei2024aniportrait} & 36.965 & 471.452 & 1.095 & 12.461 & 3.161 & 0.9508 & 0.9921 & 44.03\\
             &  V-Express~(2024)~\cite{wang2024v}   & 47.396 & 758.023 & 1.256 & 12.394 & 2.634 & 0.9093 &  \underline{0.9968} & 39.04\\
             &  Hallo~(2024)~\cite{xu2024hallo}       &  \underline{30.176} &  347.358 &  4.060 &  9.551  & 1.792 &  0.9555 & 0.9941 & 74.65\\
             &  Hallo2~(2024)~\cite{cui2024hallo2}       & 38.673 & \underline{328.540} & \underline{4.136} &  \underline{9.465}  & 2.203 & \textbf{0.9606} & 0.9942 & 45.75\\
             &  EchoMimic~(2024)~\cite{chen2024echomimic}   & 33.207 & 384.304 & 2.514 & 10.743 & \textbf{1.486} & 0.9527 & 0.9934 & 5.45\\
             & \textbf{Sonic (Ours)}       & \textbf{29.104} & \textbf{301.173} & \textbf{4.197} & \textbf{9.371} & \underline{1.745} & \underline{0.9595} & \textbf{0.9970} & 17.04\\
        \midrule
            \multirow{7}{*}{\makecell{CelebV-HQ}} 
             &  SadTalker$^*$~(2024)~\cite{zhang2023sadtalker}   & 57.574 & 841.962 & 1.978 & 10.915 & 2.252 & 0.9434 & 0.9959 & 3.75\\
             &  Aniportrait~(2024)~\cite{wei2024aniportrait} & 53.746 & 590.373 & 0.996 & 12.084 & 3.296 & 0.9522 & 0.9911 & 44.03\\
             &  V-Express~(2024)~\cite{wang2024v}   & 65.400 & 889.985 & 0.809 & 13.255 & 2.713 & 0.9019 & \underline{0.9965} & 39.04\\
             &  Hallo~(2024)~\cite{xu2024hallo}       & \underline{47.403} & 488.499 &  \underline{2.680} &  \underline{10.292}  & 2.273 & 0.9607 & 0.9942 & 74.65\\
             &  Hallo2~(2024)~\cite{cui2024hallo2}       & 52.396 & \textbf{481.336} &  2.638 &  10.343  & 2.819 & 0.9587 & 0.9942 & 45.75\\
             &  EchoMimic~(2024)~\cite{chen2024echomimic}   & 48.267 & 596.870 & 1.949 & 10.754 & \underline{2.136} & \underline{0.9612} & 0.9932 & 5.45\\
             & \textbf{Sonic (Ours)}       & \textbf{43.137} & \underline{483.108} & \textbf{2.689} & \textbf{10.194} & \textbf{1.783} & \textbf{0.9624} & \textbf{0.9972} & 17.04\\
        \toprule
\end{tabular}

    }
\end{table*}
\begin{figure*}[!ht]
    \centering
    \includegraphics[width=0.9\linewidth]{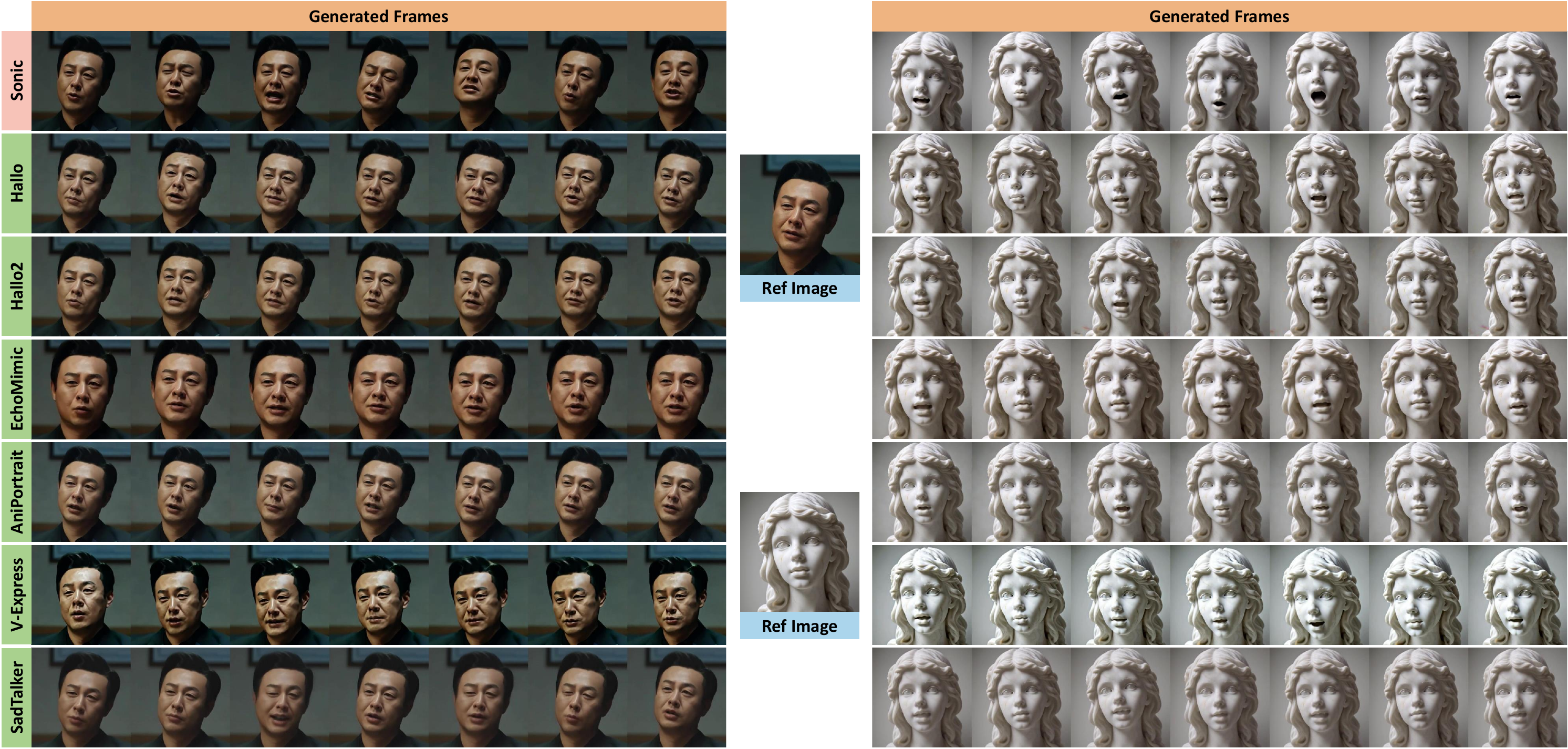} %
    \vspace{-3mm}
    \caption{\textbf{Qualitative comparisons with State-of-the-Art talking head generation methods.}
    Due to image does not reflect important sync, naturalness and stability, the full video comparison will be included in supplementary materials as well as comparison with demos from other non-open source works.} 
    \label{fig:quantitalive_compare}
    \vspace{-3mm}
\end{figure*}

\noindent \textbf{Computation cost.}
 We analyze the computation efficiency of our time-aware position shift fusion compared to motion frames and overlapping. Denote the FLOPs of a single forward pass of the video diffusion model as $\Omega$, the number of frames to generate as $n \times f$, and the denoising steps as $T$. The accumulated FLOPs of Sonic is:
\begin{equation}
\small
    \Omega(Shift) = \Omega \times T \times n.
\end{equation}
For overlapping, the overlapped latents (overlap $o$) are calculated twice,
\begin{equation}
\small
    \Omega(Overlap) = \Omega \times  T \times (n + \frac{o}{f} ) .
\end{equation}
For motion frames, the FLOPs of a single forward of Reference Net is $\omega_{r}$, the number of motion frames is $o$ and the Flops of motion modules computing $f$ frames is $\omega_{m}$,
\begin{equation}
\small
    \Omega(MF) = \Omega \times  T \times n + \omega_r \times o \times n  + \omega_m \times T \times n \times \frac{2of+o^2}{f^2}.
\end{equation}
Overall, the proposed time-aware position shift fusion significantly reduces the computation complexity.

\begin{figure*}[ht!]
    \centering
    \includegraphics[width=1\linewidth]{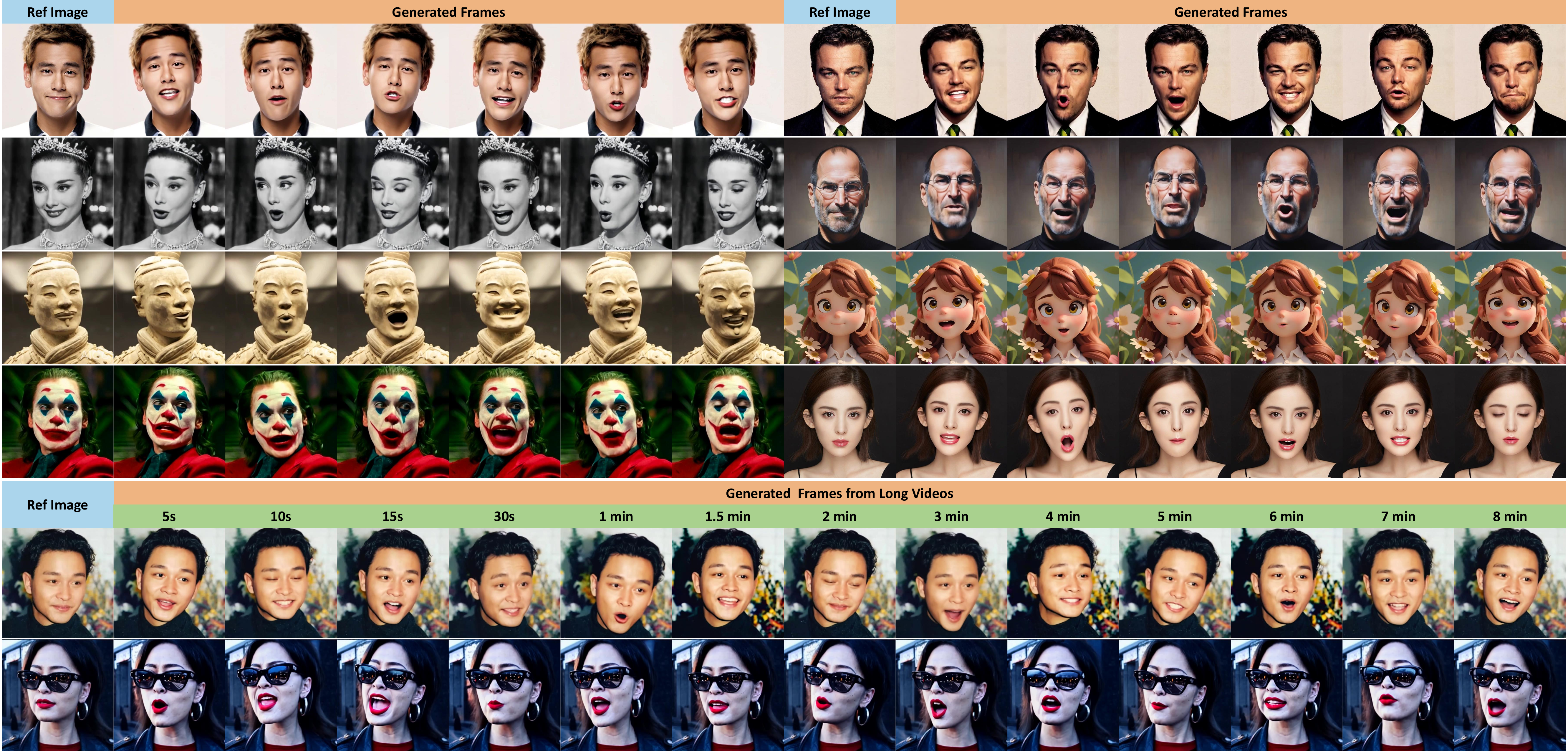} %
    \vspace{-6mm}
    \caption{\textbf{Qualitative results under different styles of portrait images and various types of audio inputs.} The images and audios were collected from recent works and the Internet. The upper section presents results for audio input ranging from $20$ seconds to $2$ minutes in duration, while the lower section shows results for longer audio inputs, up to $10$ minutes. 
    Our Sonic demonstrates versatility across various portrait styles and maintains vividness over extended durations. } 
    \label{fig:fig_quantitalive_show}
    \vspace{-3mm}
\end{figure*}

\section{Experiments}

\subsection{Experiments Setup}
\noindent \textbf{Implementation Details.}
The Sonic training is in a single stage, with initialization of the spatial module and temporal modules from stable-video-diffusion-xt-1-1\cite{blattmann2023stable}.
To separately enable different conditions in training,  we manipulate the data such that 5\% of it drops audio, another 5\% drops image, and 5\% drops both condition. 
In inference phase, multi-condition CFG is performed. The CFG of the reference image $r_i$ is set to $2.0$, and the CFG of the audio $r_a$ is set to $7.5$. In our full model, the dynamic scale $\beta$ is set to $1.0$ and the shift offset $\alpha$ is set to $7$.

\noindent \textbf{Datasets and Evaluation Metrics.}
For the training set, we created a diverse dataset by aggregating data from open-source datasets, including VFHQ \cite{xie2022vfhq}, CelebV-Text \cite{yu2023celebv}, VoxCeleb2 \cite{chung2018voxceleb2}.
For the test set, we conducted experiments on two widely used datasets, HDTF \cite{zhu2022celebv} and CelebV-HQ \cite{zhang2021flow}. We randomly selected $50$ four-second clips from CelebV-HQ and $100$ from HDTF.
We utilize seven metrics to evaluate. The Fr$\acute{\mathrm{e}}$chet Inception Distance (FID) \cite{heusel2017gans} and Fr$\acute{\mathrm{e}}$chet Video Distance (FVD) \cite{unterthiner2019fvd} metrics measure the quality of the generated data.
Expression-FID (E-FID) \cite{tian2024emo} uses face reconstruction method \cite{deng2019accurate} to extract features and calculate the FID between generated and ground truth frames, while F-SIM measures facial similarity.
SyncNet (Sync-C$\&$Sync-D) \cite{chung2017out} and VBench's smooth metric \cite{huang2024vbench} assess lip synchronization and motion fluidity. In ablation experiments, we employ VBench's dynamic degree metric to evaluate amplitude of head movements.
\subsection{Results and Analysis}
\noindent \textbf{Quantitative Comparison. }In Table \ref{tab:qualitative_experiments}, our Sonic are significantly superior than other recent 6 methods in terms of FID, Sync-confidence and Smoothness evaluated on the CelebV-HQ and HDTF datasets. 
On the HDTF dataset, our Sonic achieve the lowest FID and FVD scores, $8\%$ lower than the second-best, indicating superior generation quality and alignment with real data. 
For the CelebV-HQ dataset, which features more complex characters and backgrounds, our method maintains superior performance across most metrics and competitive FVD. We notably excel in E-FID, $16.5\%$ lower than the second-best, which demonstrates our method has superior expression diversity. 
EMO~\cite{tian2024emo} as a representative work that tackles talking head generation task, remains closed source, and not available for fair quantitative comparison. We provide video results using input from their online demos for a qualitative comparison.

\noindent \textbf{Qualitative Comparison.}
Figure \ref{fig:quantitalive_compare} provides a visualization comparison on open datasets. Analyzing the results, Previous approaches relied on motion frames, limiting the richness of generated expressions and the dynamism of head movements.
In contrast, our proposed framework incorporates global audio perception, enhancing the integration of audio and temporal variables while providing a deeper understanding of the audio input. This enables our method to generate a broader range of expressions that align with the audio and facilitate more natural head movements.

\noindent \textbf{Visualization Results in Complex Scenarios.}
We further explore the generation performance of our method under different input conditions. In terms of images, we employ three styles: real human, animation, and 3D. Each character portrait is paired with various types of audio relevant to its theme, such as singing, speech, and rap. As shown in Figure \ref{fig:fig_quantitalive_show}, our method successfully generates a wide array of expressions and diverse movements across various categories of input conditions, demonstrating strong robustness.

\begin{table}[ht!]
	\caption{User study comparison on open dataset.}
	\vspace{-3mm}
	\label{tab:user_study}
	\centering%
	\resizebox{0.99\linewidth}{!}{  
	\begin{tabular}{lcccc}
\toprule
\makecell[c]{ \textbf{Method / Metric}}   & \multicolumn{1}{c}{\textbf{Lip sync}} & \multicolumn{1}{c}{\textbf{Motion diversity}} & \multicolumn{1}{c}{\textbf{ID consistency}} & \multicolumn{1}{c}{\textbf{Video Smoothness}} \\ \hline
Aniportrait & 1.42                        & 1.62                      & 3.11                        & 2.09                          \\
SadTalker       & 1.98                        & 2.34                       & 2.95                        & 2.95                                \\
Echomimic       & 2.77                        & 2.65                       & 3.48                        & 2.71                                 \\ 
Hallo2       & 3.15                        & 2.37                       & 3.34                        & 2.94                                   \\ 
\textbf{Sonic(Ours)}       & \textbf{4.58    (45\%$\uparrow$)}                     & \textbf{4.55   (72\%$\uparrow$)}                     & \textbf{4.29    (23\%$\uparrow$)}                     & \textbf{4.66                            (58\%$\uparrow$)}       \\ 
\bottomrule
\end{tabular}

	}
\end{table}
\noindent \textbf{User Studies}
We conducted subjective evaluation on open dataset to assess the comparative methods along four key dimensions: Lip sync, motion diversity, identity similarity, and video smoothness. Total $40$ participants gives scores from 1 to 5 for results of five comparative methods. In Table~\ref{tab:user_study}, our Sonic outperforms the others in all four dimensions, especially showing a significant $72\%$ improvement in motion diversity. The video results for user study will be included in the supplementary materials.

\begin{table}[ht!]
    \caption{Quantitative results of ablations on CelebV-HQ along with our results obtained using different dynamic scales. In the Sonic model, the default setting for $\beta$ is $1.0$, \textit{i.e.} Ours-Moderate.}  
    \vspace{-3mm}
    \label{tab:ablations}
    \centering
    \resizebox{0.95\linewidth}{!}{
    \begin{tabular}{l c c c c c}
    \toprule
        \textbf{Method / Metric} & \textbf{Sync-C$\uparrow$} & \textbf{Sync-D$\downarrow$} & \textbf{E-FID$\downarrow$} & \textbf{Smooth$\uparrow$} & \textbf{Dynamic$\uparrow$} \\
    \midrule

             w/o temporal-audio & 2.610 & 10.310 & 2.311 & 0.9969 & 0.62 \\
             w/o motion-controller & 2.193 & 10.731 & 3.231 & 0.9969 & 0.22 \\
             w/o shift-fusion  & 2.509 & 10.379 & 2.202 & 0.9969 & 0.76\\ 
        \midrule

             Ours-Mild ($\beta = 0.5$)  & 2.655 & 10.242 & 2.060 & \underline{0.9970} & 0.78\\
             Ours-Intense ($\beta = 2.0$) & \textbf{2.803} & \textbf{10.125} & \underline{2.038} & 0.9969 &  \textbf{0.98} \\
          Ours-Moderate ($\beta = 1.0$) & \underline{2.689} & \underline{10.194} & \textbf{1.783} & \textbf{0.9972} & \underline{0.86}\\

    \bottomrule
\end{tabular}

    }
    \vspace{-3mm}
\end{table}

\begin{figure*}[ht!]
    \centering
    \includegraphics[width=1\linewidth]{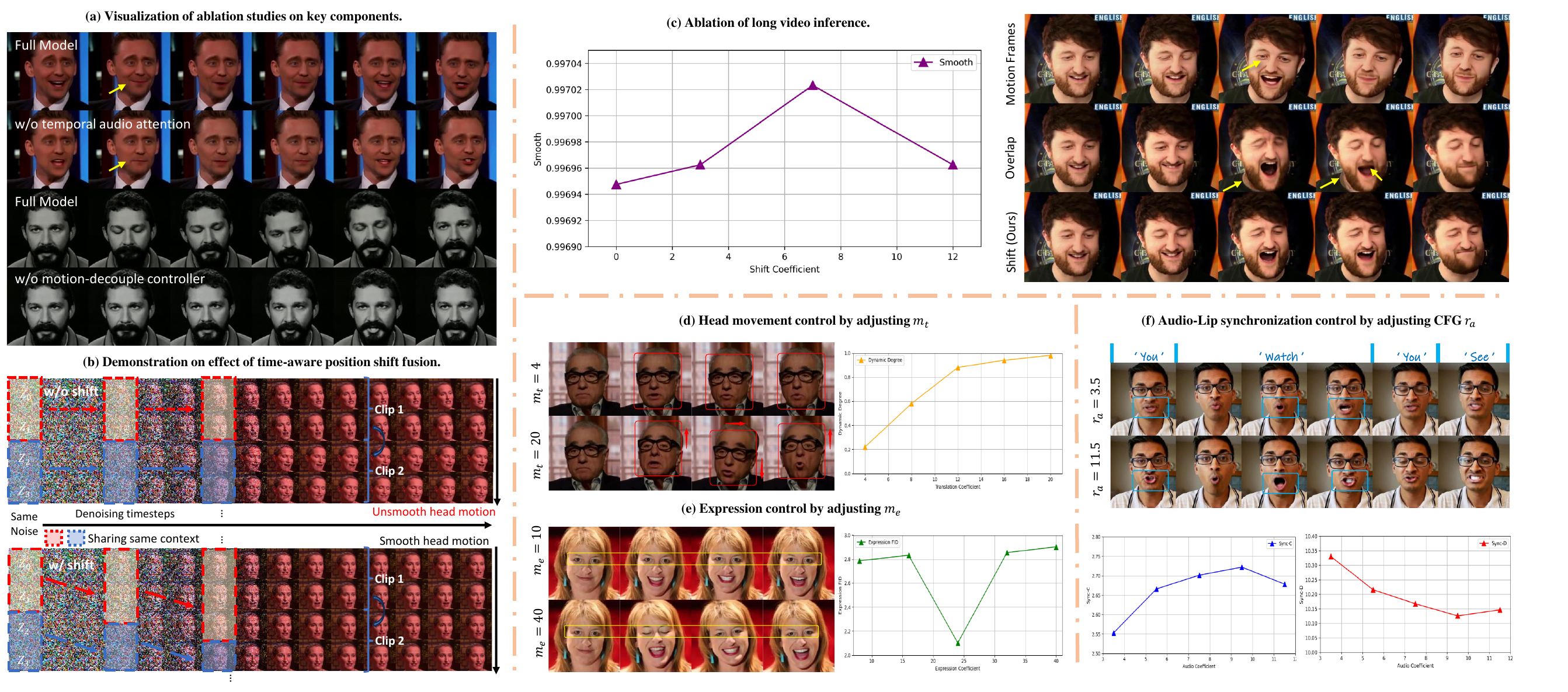} %
    \vspace{-8mm}
    \caption{\textbf{Ablation study of intra-clip and inter-clip components and adjustable parameters}. \textbf{(a)}: Visual results for ablation on temporal audio attention and motion controller. \textbf{(b)}: Demonstration for time-aware shift fusion. \textbf{(c)}: Visual results compared with motion frames and overlapping. \textbf{(d-f)}: Adjustability over lip, expression, and head movements. 
    } %
    \label{fig:fig_ablation_all}
\end{figure*}

\begin{table}[ht!]
    \caption{Quantitative comparison on motion frames, overlap and the proposed time-aware shift fusion.}  
    \vspace{-3mm}
    \label{tab:long_video_inference}
    \centering
    \resizebox{0.95\linewidth}{!}{
    \begin{tabular}{@{}l c c c c @{}}
\toprule
\textbf{Method / Metric} & \multicolumn{1}{l}{\textbf{Sync-C$\uparrow$}} & \multicolumn{1}{l}{\textbf{Sync-D$\downarrow$}} & \multicolumn{1}{l}{\textbf{Smooth$\uparrow$}} & \textbf{Runtime$\downarrow$}  \\ \midrule
Motion frames(8)     & 2.465                      & 10.323                     & 0.9969                     &    19.70   \\
Overlap(8)           & 2.544                      & 10.265                     & 0.9969                     & 22.49       \\
Shift-fusion             & \textbf{2.689}                      & \textbf{10.194}                     & \textbf{0.9972}                     & \textbf{17.04}               \\ \bottomrule
\end{tabular}

    }
    \vspace{-5mm}
\end{table}

\subsection{Ablation Studies}
\noindent \textbf{Temporal Control Inspires Intra-clip Diversity.}
In Table~\ref{tab:ablations}, the absence of temporal audio attention results in a decrease in both lip-sync and E-FID. We observe that the motion controller plays a crucial role in CelebV-HQ given that its videos exhibit more movement that weakly correlates with audio.
We also investigate the effect of different dynamic scale. Increasing $\beta$ can further stimulate the lip-audio synchronization and the diversity of movements, while there is a slight decrease in stability and expression similarity.
The instances in Figure~\ref{fig:fig_ablation_all} (a) indicates temporal audio attention helps learn effective control of micro-expressions, while the motion controller is critical for agility and expressiveness.

\noindent \textbf{Larger Audio Perception Bridges Inter-clip Stability. }
Initially, we ablate the time-aware position shift fusion (briefly termed Shift) and generate clips independently as illustrated in Figure~\ref{fig:fig_ablation_all} (b). The fusion process is performed at each timestep, gradually inverting the random noise to continuous and smooth long range videos.
Subsequently, we explore different long video inference strategy: motion frames, overlap, our shift. The number of motion frames and overlap frames was set to $8$. 
As indicated in Table~\ref{tab:long_video_inference}, our shift approach attains superior lip-sync and smoothness without necessitating any auxiliary computation. 
From the qualitative comparison demonstrated in Figure~\ref{fig:fig_ablation_all} (c), our method successfully handle long term consistency, circumventing identity similarity decrease in motion frames and mixed blurry texture in overlapping results. The left curves describe the effect of $\alpha$ on smoothness, where $\alpha=7$ yields the smoothest results.

\noindent \textbf{Fine-grained Control over Lip, Expression and Motion.}
Figure \ref{fig:fig_ablation_all} (d-f)  illustrates the varied outcomes of three adjustable parameters that facilitate fine-grained control, $m_t$ for translation, $m_e$ for expression, and audio CFG $r_a$ for lip. 
The findings in (d) reveals that larger $m_t$ significantly enhance the range of motion and dynamic degree, while smaller $m_t$ restrict movement mainly to the mouth. 
We found that increasing $m_e$ leads to more exaggerated facial expressions, though the E-FID metric does not improve, as moderate parameters align better with real data, while extreme values deviate, as shown in (e). 
(f) shows that higher $r_a$ improve lip-audio synchronization, but excessively high values do not always yield better outcomes. In our final model, the $r_a$ was set to $7.5$ to achieve an optimal balance between video quality and audio synchronization.

\section{Conclusion}
In summary, our work present Sonic, an audio-driven portrait animation that focus on the exploration of global audio perception to enhance the efficient generation of realistic lip synchronization, naturalness and temporal consistencies. The technical cornerstone mainly lies on context-enhanced audio learning and time-aware position shift fusion that work together to extent the intra-clip audio temporal to the global inter-clip audio perception. It significantly outperforms existing SOTA methods in video quality, motion diversity and naturalness.

\section{Contributors}
\noindent \textbf{Tencent Hunyuan:} 
 Linqing Wang, Zixiang Zhou, Zhentao Yu, Hongmei Wang, Yuan Zhou, Joeyu Wang, Chengfei Cai, Shiyu Tang, Tianxiang Zheng, Junshu Tang, Shuai Shao
 
\noindent \textbf{Tencent Music Lyra Lab:} 
Yubin Zeng, Junxin Huang, Yue Zhang, Chao Zhan, Zhengyan Tong, Zhaokang Chen, Bin Wu, Wenjiang Zhou

\noindent \textbf{Tencent AMS:} 
 Yuang Zhang, Junqi Cheng, Jiaxi Gu, Fangyuan Zou

{
    \small
    \bibliographystyle{ieeenat_fullname}
    \bibliography{main}
}

\end{document}